\documentclass[%
 aip,
 amsmath,amssymb,
 reprint,%
]{revtex4-1}
\usepackage[T1]{fontenc}
\usepackage[utf8]{inputenc}
\usepackage{xcolor}
\usepackage{bm}
\usepackage{amsmath}
\usepackage{amssymb}
\usepackage{graphicx}
\usepackage{ulem}

\usepackage[unicode=true,
 bookmarks=false,
 breaklinks=false,pdfborder={0 0 1},backref=section,colorlinks=false]
 {hyperref}
\usepackage{breakurl}

\makeatletter

\providecommand{\tabularnewline}{\\}
\usepackage{tikz}
\usetikzlibrary{calc}

\usepackage{dcolumn}
\usepackage{bm}
\usepackage{amsthm}
\usepackage{amsfonts}
\usepackage{float}
\usepackage{color}
 \usepackage{tcolorbox}
\usepackage{tabularx}
\usepackage{enumitem}
\setlist[itemize]{leftmargin=*}

\DeclareMathOperator{\sign}{sign}
\DeclareMathOperator{\diag}{diag}
\DeclareMathOperator{\arctan2}{arctan2}

\makeatother

\begin{document}

\title{Deep Time-Delay Reservoir Computing: Dynamics and Memory Capacity}

\date{\today}

\author{Mirko Goldmann}
\altaffiliation[Electronic mail: ]{mirko-goldmann@hotmail.de}
\affiliation{Institute of Theoretical Physics, Technische Universität Berlin,
D-10623, Germany}
\affiliation{Department of Mathematics, Humboldt-Universität zu Berlin,
D-12489, Germany}

\author{Felix Köster}

\affiliation{Institute of Theoretical Physics, Technische Universität Berlin,
D-10623, Germany}

\author{Kathy Lüdge}

\affiliation{Institute of Theoretical Physics, Technische Universität Berlin,
D-10623, Germany}

\author{Serhiy Yanchuk}

\affiliation{Institute of Mathematics, Technische Universität Berlin, D-10623,
Germany}
\begin{abstract}
The Deep Time-Delay Reservoir Computing concept utilizes unidirectionally connected systems with time-delays for supervised learning. We present how the dynamical properties of a deep Ikeda-based reservoir are related to its memory capacity (MC) and how that can be used for optimization. In particular,
we analyze bifurcations of the corresponding autonomous system
and compute conditional Lyapunov exponents, which measure the generalized
synchronization between the input and the layer dynamics. We show
how the MC is related to the systems distance to bifurcations
or magnitude of the conditional Lyapunov exponent. The interplay of different
dynamical regimes leads to an adjustable distribution between linear and 
nonlinear MC. Furthermore, numerical simulations show resonances between
clock cycle and delays of the layers in all degrees of the MC. Contrary
to MC losses in single-layer reservoirs, these resonances can boost
separate degrees of the MC and can be used, e.g. to design a system with maximum linear MC. Accordingly, we present two configurations that empower either high nonlinear MC or long time linear MC.
\end{abstract}
\maketitle
\begin{quotation}
The brain-inspired reservoir computing paradigm manifests the
natural computing abilities of dynamical systems. Inspired by randomly
connected artificial neural networks called echo state networks, simple
optic and opto-electronic hardware implementations were developed,
opening the research for delay-based reservoirs. These systems show promising
performance at different supervised machine learning tasks like time-series
forecasting, e.g., for the chaotic Mackey-Glass attractor, but also
in speech recognition. The implementations employ a dynamical node with delayed feedback, which can exhibit multi-dimensional complex dynamics. In
delay-based reservoir computing, the nodes of the network are separated temporally,
and the computation time correlates with the number of nodes. An ongoing
search for systems with improved performance started, resulting in
more complex implementations. In this paper, we analyse the concept
of deep time-delay reservoir computing, where multiple delay
systems, called layers, are coupled unidirectionally. Such a scheme
enables a constant low computation time while the number of nodes
increases via additional layers. We investigate the dynamics of the layers and explain the effects of their interplay, where the influence
onto the computational capabilities are measured using the linear and nonlinear memory. By utilizing this interplay, we show a strong adaptability
of the reservoirs performance and we show ways to optimize, e.g. the linear memory of a reservoir computer. 
\end{quotation}

\section{Introduction}

The introduction of the reservoir computing paradigm by Jaeger \cite{Jaeger2001} and
Maass \cite{Maass2002} independently gained considerable interest
in supervised machine learning utilizing dynamical systems. The reservoir computing
scheme contains three different parts: the input layer, the reservoir,
and the output layer. The reservoir can be any dynamical system, like an artificial neural network but also a laser with self-feedback. The output layer
is trained by a linear weighting of all accessible reservoir states,
while the reservoir parameters are kept fixed. This simplification overcomes
the main issues of the time expensive training of recurrent neural
networks like exploding gradients and its high power consumption \cite{Werbos1990}.

Appeltant et al.~\cite{Appeltant2011} successfully implemented the
RC scheme onto a single nonlinear node with a delayed self-feedback.
In the input layer, time-multiplexing is used to create temporally
separated virtual nodes. The reservoir dynamics are hereby given by
a delay differential equation, which has been proven to exhibit
rich high-dimensional dynamics \cite{Hale1993,Erneux2009,Erneux2017,Yanchuk2017}.
For the training, the temporally separated nodes are read out and
weighted to solve a given task. The introduction of time-delay reservoir computing enabled simple optical and opto-electronic hardware implementations,
which led to improvements of computation time scales for supervised
learning \cite{Brunner2018,Larger2017}. The delay-based reservoirs were successfully
applied to a wide range of tasks, such as  chaotic time series forecasting
or speech recognition.

The success of single node delay-based reservoir computing has triggered
interest into more complex network architectures, like coupled Stuart-Landau oscillators arranged in a ring topology~\cite{Rohm2018},
single nonlinear nodes with multiple self-feedback loops of various
length \cite{Chen2019}, parallel usage of multiple nodes~\cite{Sugano2020}
and multiple nonlinear nodes coupled in a chain topology \cite{Penkovsky2019a}.
Further, it was recently shown by Gallicchio et al.~\cite{Gallicchio2018,Gallicchio2019,Gallicchio2017c},
that echo state networks with multiple unidirectional coupled layers
called deepESN provide a performance boost in comparison to their
shallow counterpart. 
Various cascading reservoir setups were studied in 
\cite{Keuninckx2017,Freiberger2020}.
Penkovsky et al.~\cite{Penkovsky2019a} found
that unidirectional coupled delay systems are superior against bidirectional
coupling for certain symmetric parameter choice.

In the following,
we present a deep time-delay reservoir computing model, where
we use asymmetrical layers and discuss their dynamical and computational
properties. 
In contrast to the general deepESN scheme, the considered model 
possesses the same number of nodes in all layers, which is the result of our time multiplexing 
procedure for constructing a network from a time-delay system. 
Our setup is, in a certain sense, more simple than the cascading reservoirs considered in
\cite{Keuninckx2017,Freiberger2020}, since no output signals (e.g., linear regression) are generated at each layer separately. The input enters only the first layer, and each consecutive layer receives only the dynamical state of the previous layer. 
As a result, we do not perform sequential training of the layers.

The paper is structured as follows: In section \ref{deepTRC} we present
the system implementing deep time-delay reservoir computer and 
show its performance at predicting the chaotic Mackey-Glass attractor.
Afterward in \ref{Dynamics}, we study the dynamical properties
of an autonomous two-layer system. The conditional Lyapunov exponent
for a non-autonomous system is introduced in \ref{CLE}. The numerically
calculated conditional Lyapunov exponent is then related to the linear and nonlinear memory
capacity. In section \ref{DelayClockTimeResonances} the resonances
of the clock cycle and the delay-times are presented for two and three-layer
systems. 

\section{Deep Time-Delay Reservoir Computing\label{deepTRC}}

\subsection{Model}

A deep time-delay reservoir computer (deepTRC) consists of $L\in\mathbb{N}$
nonlinear nodes with states $\boldsymbol{v}_{l}(t)\text{ for }l=1,2,\dots,L$.
All $L$ nodes feature a self-feedback with a delay length $\tau_{l}>0$.
\begin{figure}[h]
\centering \includegraphics[width=0.95\linewidth]{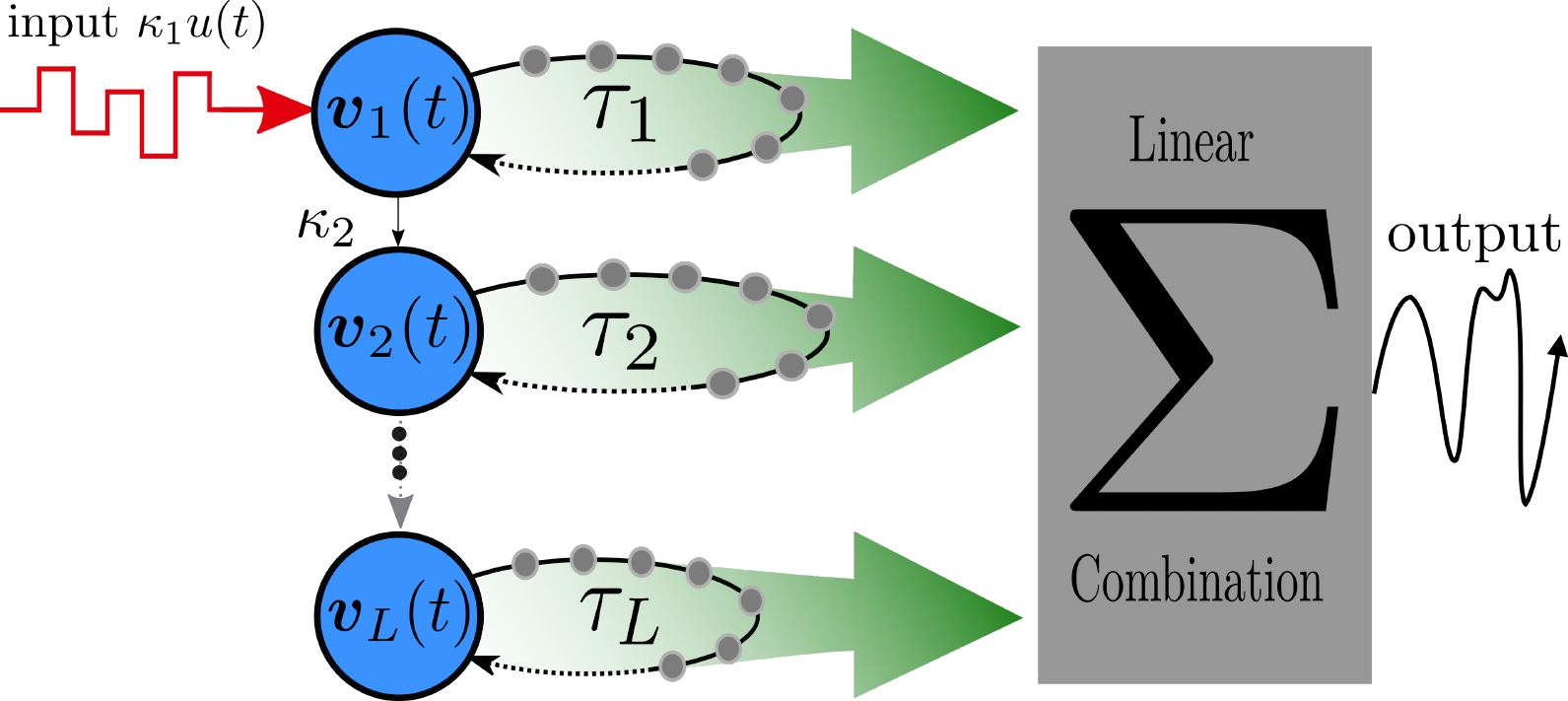} \caption{Deep Time-Delay Reservoir Computer containing $L$ layers (blue) with
delayed self-feedback, where the delay length $\tau_{l}$ can vary.
The first layer $\boldsymbol{v}_{1}(t)$ is driven by the time multiplexed
input sequence $\kappa_{1}u(t)$ whereas all other layers are driven
by their previous layer weighted with $\kappa_{l}$. The output layer
is given by linear weighting of all states.}
\end{figure}

The nodes are coupled unidirectional, where the first node is the
only which is feed by the task-specific input sequence $u(t)$. The
coupling between the nodes is instantaneous, i.e., without delay. The
nodes with their corresponding feedback loops are referred to as layers
in the following because of the unidirectional topology and their
high-dimensional dynamics. The dynamical evolution of a layer $l$
is given by: 
\begin{equation}
\dot{\boldsymbol{v}}_{l}(t)=F_{l}(\boldsymbol{v}_{l}(t),\boldsymbol{v}_{l}(t-\tau_{l}),J_{l}(t))
\end{equation}
with $F_{l}(\dots)$ being a nonlinear delay differential equation, and $J_{l}(t)$ the layer
dependent input: 
\begin{equation}
J_{l}(t)=\begin{cases}
u(t)\text{ for }l=1,\\
v_{l-1}^1(t)\text{ else.}
\end{cases}
\end{equation}
In the following, we assume that the coupling to the layer $l$ is realized via the first component $x_{l-1}=v_{l-1}^1(t)$ of the dynamical variable $\boldsymbol{v}_{l-1}(t)$ of layer $l-1$. Time-multiplexing is used to transform the layers into high dimensional
networks. The discrete input is given by the sequence $(s(k))_{k\in\mathbb{N}_{0}}$,
$s(k)\in\mathbb{R}$. This sequence is transformed into the time-continuous
input $u(t)$ as follows 
\begin{equation}
u(t)=u_{k,j}=s(k)m_{j}\text{ for }t\in[kT+(j-1)\theta,kT+j\theta],\label{eq:timemultiplexing}
\end{equation}
where $T$ is the clock cycle and the scaling $m_{j}$, $j=1,\dots N_{V}=T/\theta$
determines a mask, which is applied periodically with the period $T$.
Such a preprocessing method generates $N_{V}$ virtual nodes which
are distributed temporally with a separation of $\theta=T/N_{V}$,
see more details in \cite{Appeltant2011,Rohm2018}. The given deepTRC
now contains $L$ layers with each having $N_{V}$ virtual nodes resulting
in a total reservoir size of  $N_{R}=LN_{V}$. The virtual nodes within
the layers correspond to the values $x_{l}(kT+j\theta)=\boldsymbol{v}_{l}^1(kT+j\theta)$. 

For the training of the deepTRC, all virtual nodes $x_{l}(kT+j\theta),j=1,\dots,N_{V}$
of each layer $l=1,\dots,L$ are read out. The virtual nodes are combined
into the global state $X(k)\in\mathbb{R}^{N_{R}}$ given by
\begin{align}
X(k):=\left(\begin{matrix}X_{1}\\
X_{2}\\
\vdots\\
X_{N_{R}}
\end{matrix}\right):=\left(\begin{matrix}x_{1}(kT)\\
x_{1}(kT+\theta)\\
\vdots\\
x_{L}(kT+(N_{V}-2)\theta)\\
x_{L}(kT+(N_{V}-1)\theta)
\end{matrix}\right).
\end{align}
In order to train for a given task $\hat{o}(k)$, the global state
is weighted    
\begin{align}
o(k) & =W^{T}X(k)+c,
\end{align}
where $c$ is a constant bias and the weights $W\in\mathbb{R}^{N_{R}}$
are determined via a linear regression with an optional Tikhonov regularization.

In the following, we will focus on the analysis of the recently introduced
opto-electronic reservoir~\cite{Argyris2020,Penkovsky2019a,Larger2012a,Soriano2013,Soriano2015,VanDerSande2017a,Brunner2018,Chen2019}, which is governed by the
equations: 
\begin{eqnarray}
\dot{x}_{l}(t) & = & -x_{l}(t)-\delta_{l}y_{l}(t)\nonumber \\
 & + & \beta_{l}\sin^{2}(x_{l}(t-\tau_{l})+\kappa_{l}J_{l}(t)+b_{l}),\label{ikedaEquations}\\
\dot{y}_{l}(t) & = & x_{l}(t),\nonumber 
\end{eqnarray}
where $\kappa_{1}$ is the input gain and $\kappa_{l},$ $l>1$ are
coupling strengths between the consecutive layers $l-1$ and $l$.
Further, $\delta_{l}$ is a damping constant satisfying $\delta_{l}\leq\delta_{l+1}$,
$\beta_{l}$ is the feedback gain and $b_{l}$ is a scalar phase shift
of the nonlinearity. The dynamical variable of a layer becomes $\boldsymbol{v}_{l}=\left(x_{l},y_{l}\right)^{T}$. According to the $\sin^{2}$ nonlinearity, the system is referred as Ikeda time-delay system \cite{Penkovsky2019a,Gao2019,Ikeda1979}. 

\subsection{Chaotic Time-Series Prediction}

\begin{figure}[b]
\centering \includegraphics[width=1\linewidth]{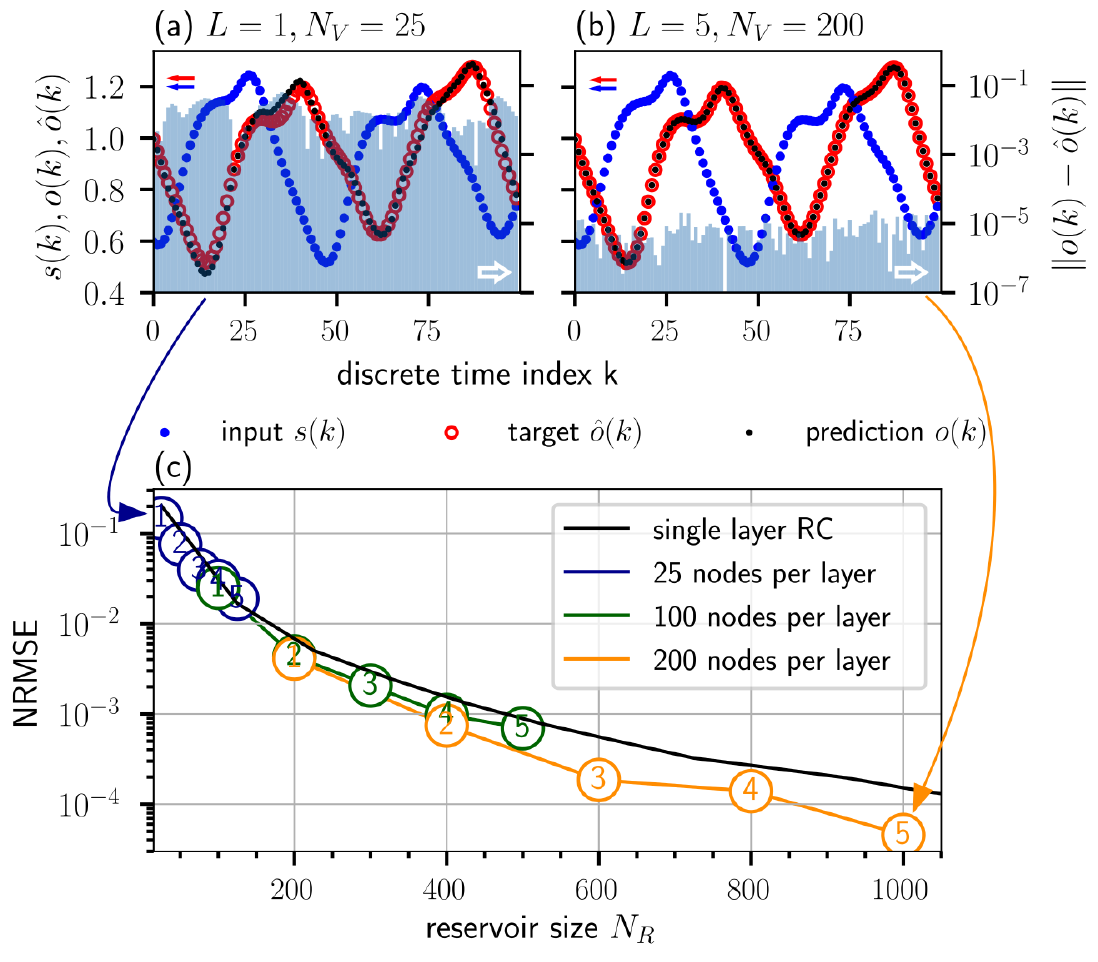}
\caption{\label{performance} Input $s(k)$, output $\hat{o}(k)$ and prediction $o(k)$ of the chaotic Mackey-Glass task of (a) a single-layer system with 25 virtual nodes ($N_R=25$) and (b) a five-layer system with 200 virtual nodes per layer ($N_R=1000$).
Hereby the output $\hat{o}(k)$ is a temporal shift of the input sequence $\hat{o}(k) = s(k+34)$. (c) Performance comparison of deepTRCs with up to $L=5$ layers at the Mackey-Glass prediction task $\Delta n=34$ into future.
The black line represents the performance of a single-layer TRC with the increase of virtual nodes $N_{V}$. The colored lines represent deepTRC, where the number in the circle marks the number of layers, e.g. orange circle with 3 means 3 layers with each having 200 nodes. For all shown simulations the separation of nodes was kept fixed at $\theta=1$. Accordingly the clock cycle $T=N_V \theta$ increases with the amount of virtual nodes $N_V$.}
\end{figure}

In Fig.~\ref{performance}, we show the performance of the Ikeda
based deepTRC at the prediction of the chaotic Mackey-Glass attractor,
i.e., $\hat{o}(k)=s(k+\Delta n),$ where $s(t)$ is the time-series
of the Mackey-Glass system and $\Delta n$ determines how far into
the future it shall be predicted. The chosen prediction step $\Delta n=34$
 corresponds to twice the delay time of the Mackey-Glass system which is shown in Fig.~\ref{performance}~(a) and (b). The parameters of the Mackey-Glass
system are as in Jaeger et al.~\cite{Jaeger2004a}. We simulated
deepTRC with up to $L=5$ layers and varied the number of nodes per layer. Hereby, the separation of nodes is set to $\theta=1$ and will be kept fixed for all following simulations. Accordingly the clock cycle $T$ increases as the number of virtual nodes $N_V$ increases.
A Bayesian optimization approach~\cite{Snoek2012} was used to optimize
the feedback-gains $\beta_{l}$, the delays $\tau_{l}$, coupling gains $\kappa_l$ and phase offsets $b_l$. Hereby,
the Bayesian optimisation of deepTRC becomes much harder for deeper systems
according to an increased amount of hyperparameters. In general, additional
layers can improve the performance of deepTRC as compared to single-layer
TRC as it is shown in (a) and (b). As a remark, the deepTRC enables a faster computation by a constant separation of nodes $\theta$ , i.e a deepTRC with $L=5$ layers is $5$ times faster than the single-layer TRC with the same amount of total nodes in line with a 5 times shorter clock
cycle. For the evaluation of the performance an initial input of length $K_{\text{init}}$ = $10^{4}$, a training length of $K_{\text{train}}$ = $10^{4}$ and a testing length of $K_{\text{test}}$ = $2000$ is used.

The parameters of the best deepTRC with $L=5$ layers are shown in table \ref{tab:performance}. 
Note that the coupling gains of the last three layers $\kappa_i$ are small. This  
indicates that these layers might play the role of linear filtering of the input signal, with  an additional mixing due to different delays. We discuss the role of the layers detailed in Sec.~\ref{DeepSpectrum}.
\begin{table}
    \centering
    \begin{tabular}{|c|c|c|c|c|c|}
        \hline
          & delay $\tau$ & damping $\delta$ & phase $b$ & feedback  $\beta$ & coupling $\kappa$\\ \hline \hline
         Layer 1 & 230 & 0    & 0.2  & 0.68 & 4.0 \\  \hline
         Layer 2 & 457 & 0.01 & 0.2  & 0.8  & 1\\  \hline
         Layer 3 & 199 & 0.01 & 1.5 &  0.97 & 0.01\\ \hline 
         Layer 4 & 27  & 0.01 & 1.28 & 0.83 & 0.13\\ \hline 
         Layer 5 & 40  & 0.01 & 1.9 & 0.2 & 0.01\\ \hline 
    \end{tabular}
    \caption{Parameters of the five-layer deepTRC with the best performance (NRMSE$=4.56\times10^{-5}$) shown in Fig.~\ref{performance}~(b). The deepTRC has $N_V=200$ virtual nodes per layer and the clock cycle is $T=200$.}
    \label{tab:performance}
\end{table}

\section{Dynamics of Autonomous deepTRC\label{Dynamics}}

The dynamics of a delay-based RC play an essential role for its performance
 \cite{Appeltant2011,Dambre2012}. In this section we consider  an
autonomous $L$-layer deepTRC by setting $J^{(1)}=0$. 

The equilibrium of Eq.~\eqref{ikedaEquations} are given as solutions
of the following nonlinear system of equations 
\begin{align}
\begin{cases}
 & x_{l}^{\ast}=\beta_{l}\sin^{2}(x_{l}^{\ast}+b_{l})\qquad\text{if }\delta_{l}=0\\[2ex]
 & \!\begin{aligned}x_{l}^{\ast} & =0,\\
y_{l}^{\ast} & =\frac{\beta_{l}}{\delta_{l}}\sin^{2}(\kappa_{l}x_{l-1}^{\ast}+b_{l})
\end{aligned}
\qquad\text{if }\delta_{l}>0
\end{cases}\label{eq:nonl-syst}
\end{align}
Without further restriction we set $\delta_{1}=0$ and $\delta_{l>1}>0$.
System (\ref{ikedaEquations}) can be linearised around this equilibrium,
which leads to:
\begin{equation}
\begin{pmatrix}
\dot{\bm{\xi}_1}(t)\\
\dot{\bm{\xi}_2}(t)\\
\vdots\\
\dot{\bm{\xi}_L}(t)
\end{pmatrix}
=A
\begin{pmatrix}
\bm{\xi}_1(t)\\
\bm{\xi}_2(t)\\
\vdots\\
\bm{\xi}_L(t)
\end{pmatrix}
+B
\begin{pmatrix}
\bm{\xi}_1(t-\tau_1)\\
\bm{\xi}_2(t-\tau_2)\\
\vdots\\
\bm{\xi}_L(t-\tau_L)
\end{pmatrix}
\label{eq:linearised}
\end{equation}
where $\bm{\xi}_l(t)$ 
is the linearisation of $l^{th}$-layers dynamical variable $\boldsymbol{v}_l(t)$ and $\bm{\xi}_l(t-\tau_l)$ 
is the linearisation shifted by the delay of the layer.

The block matrix $A\in \mathbb{R}^{m \times m}, m=2L$ is given by the sub-matrices $A_{l,l'} \in \mathbb{R}^{2\times2}$ 
\begin{equation}
	A_{l,l'} = 
	\frac{d F_l(\bm{v}_l(t),\bm{v}_l(t-\tau_l),J_l(t))}{d\bm{v}_{l'}(t)},
\end{equation}
According to the unidirectional topology of the Ikeda-deepTRC matrix $A$ becomes a lower triangular block matrix. Further, the block matrix $B \in \mathbb{R}^{m\times m}$ can be calculated by:
\begin{equation}
    B_{l,l'} =  \frac{d F_l(\bm{v}_l,\bm{v}_{l}(t-\tau_l),J_l(t)}{d\bm{v}_{l'}(t-\tau_{l'}))}.
\end{equation}
Therefore, $B=\diag(B_{11},B_{22},\dots,B_{LL})$ becomes block diagonal. The sub-matrices of the autonomous Ikeda deepTRC \eqref{ikedaEquations} are given by:
\begin{align}
    A_{1,1} &= 
    \begin{pmatrix}
        -1 & 0 \\
        0 & 0
    \end{pmatrix}, 
    \qquad A_{l,l} = 
    \begin{pmatrix}
        -1 & -\delta_l \\
        1 & 0
    \end{pmatrix} \text{ for } l>1,\nonumber\\
        \qquad A_{l+1,l} &= 
    \begin{pmatrix}
        \beta_{l+1}\kappa_{l+1}\nu_{l+1} & 0 \\
        0 & 0
    \end{pmatrix},
    \qquad
    B_{l,l} =
    \begin{pmatrix}
        \beta_l \nu_l & 0\\
        0 & 0
    \end{pmatrix}
\end{align}
with $\nu_{1}=\sin(2(x_{1}^{*}+b))$ and $\nu_{l>1}=\sin(2(\kappa_l x_{l-1}^{*}+b_{l})$.
The Eigenvalues of the linearised system can therefore be calculated by solving the characteristic equation:
\begin{align}
0=\left|A-\lambda\bm{1} + \tilde{B}\right|
\end{align}
with $\tilde{B} = \diag({B}_{1,1} \exp(\lambda \tau_1),\dots,B_{L,L}\exp(\lambda \tau_L))$.
This equation can further be simplified by using that the determinant of a lower triangular matrix is given by the product of the determinants of the block matrices:
\begin{equation}
 0 = \prod_{l=1}^{L}|A_{l,l} - \lambda \bm{1} + {B}_{l,l} \exp(\lambda \tau_1)  |
\end{equation}
The characteristic equation of our $L$-layer Ikeda deepTRC \eqref{ikedaEquations} is therefore given by:
\begin{align}
    0 =& (-\lambda-1+\beta_1\nu_1\exp(\lambda \tau_1)) \nonumber \\ &\times\prod_{l=2}^L(\delta_l+\lambda(\lambda+1-\beta_l\nu_l)\exp(\lambda \tau_l)))
\end{align}

For the details of the derivation we refer to Appendix~\ref{app:StabilityAnalysis}.
One can see that  the eigenvalues of the linearised autonomous deepTRC
are given by the combined set of the eigenvalues for the single layers.

\begin{figure*}[ht]
\centering \includegraphics[width=1\linewidth]{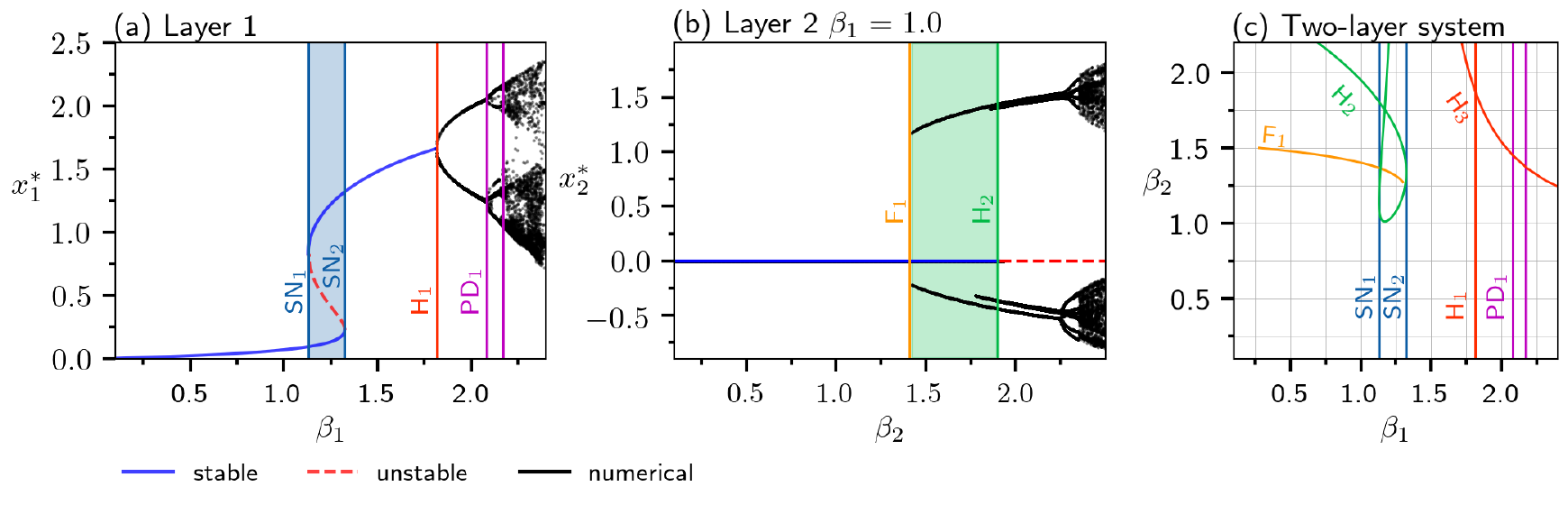}
\caption{\label{dynamics}(a) Bifurcation diagram of the first layer where the
blue shaded area marks bi-stability, (b) bifurcation diagram of the second layer for constant $\beta_1=1.0$ where the green shaded area marks coexistence of a stable equilibrium $x_2^\ast=0$ and a stable limit cycle, (c) 2-parameter bifurcation diagram
of the two-layer system. The stability and the Hopf bifurcations are computed using the analytical results presented in Appendix \ref{app:StabilityAnalysis}, all other bifurcations are computed using the DDE-Biftool. The used parameters are given in Table~\ref{tab:Parameter-values}. \textbf{Bifurcation
legend:} H (red) –  supercritical Hopf, H (green) – subcritical Hopf,
SN – saddle-node (blue),F – fold (orange) and PD – period doubling (magenta) bifurcation. For the periodic and chaotic parts of the bifurcation diagram in (a) and (b), the maxima and minima of the solutions are plotted.}
\end{figure*}

In the following we study the dynamics of the  two-layer deepRTC depending
on the feedback gains $\beta_{1}$ and $\beta_{2}$. The remaining
parameters are kept fixed to the values shown in Table~\ref{tab:Parameter-values}. 

\begin{table}
\begin{tabular}{|c|c|c|c|c|}
\hline 
Description & Parameter & Layer 1 & Layer 2 & Layer 3\tabularnewline
\hline 
\hline 
delay time & $\tau$  & 30 & 30 & 30\tabularnewline
\hline 
feedback gain & $\beta$ & 1.6 & 1.3 & 1.3\tabularnewline
\hline 
phase offset & $b$ & 0.2 & 0.2 & 0.2\tabularnewline
\hline 
coupling & $\kappa$  & 0.0 & 1 & 1\tabularnewline
\hline 
damping & $\delta$  & 0 & 0.01 & 0.01\tabularnewline
\hline 
initial value & $\bm{v}_{l,0}$  & $[0.07,0]^T$ & $[0,0]^T$ & $[0,0]^T$\tabularnewline
\hline 
\end{tabular}

\caption{\label{tab:Parameter-values} Parameter values used for the calculation of the bifurcation diagrams in Fig.~\ref{dynamics}. Additionally, the parameters are used for the conditional Lyapunov exponent in Fig. \ref{lyapExpBetaPlane} and the memory capacity in Fig. \ref{betaPlane} and \ref{fig:tauDPlane}, whereas there the input was enabled by setting the input gain $\kappa_1=0.01$.}
\end{table}

\begin{center}
\par\end{center}

In Fig.~\ref{dynamics}~(a) the equilibrium of the first layer $x_{1}^{*}$
is shown as a function of the feedback gain $\beta_{1}$. The layer
exhibits two saddle-node bifurcations at $\beta_{1}\approx1.15$ and
$\beta_{1}\approx1.3$ respectively; between these two points the
first layer possesses two coexisting stable equilibria. Further, it
reveals periodic solutions after the supercritical Hopf bifurcation
at $\beta_{1}\approx1.816$, and  a period-doubling cascade resulting
in chaotic dynamics at $\beta_{1}\approx2.2$. The numerical part
of the bifurcation diagram, i.e., the chaotic solutions of $x_{1}(t)$
in Fig.~\ref{dynamics}~(a) are computed using Heun's method. The stability and the location of the Hopf bifurcations are calculated using the derivation in Appendix \ref{app:StabilityAnalysis}.

In Fig~\ref{dynamics}~(b) the bifurcation diagram of the second layer is shown for a constant feedback gain of the first layer $\beta_1=1.0$. The second layer reveals a subcritical Hopf at $\beta_2\approx1.9$. The created unstable limit cycle becomes stable due to a saddle-node bifurcation at $\beta_2\approx1.4$. Accordingly, a coexistence of a limit cycle and the stable equilibrium $x_2^\ast=0$ occurs in the range $\beta_2\in[1.4,1.9]$.

While the first layer can be considered separately, the second layer
is driven by the first one. As a result, one has to study the whole
system (\ref{eq:2-layers}) for analyzing the dynamics of the second
layer. In Fig.~\ref{dynamics}~(c) the bifurcation diagram of the
full two-layer system  is obtained  using the DDE-Biftool \cite{Sieber2016}.
The bifurcations of the first layer occur as vertical lines (indicating saddle-node, Hopf and period-doubling bifurcations in blue, red and magenta) while the sub and supercritical Hopf and the fold bifurcations of the second layer occur as curved lines in the $\beta_1$,$\beta_2$ plane (green, red and orange). 

In the following sections, the obtained bifurcation diagrams will
be compared with the other characteristics of the reservoir that describe
its memory capacity. In particular, the next section investigates conditional
Lyapunov exponents and shows how they restrict the parameter
set where the RC can properly function.

\section{Conditional Lyapunov Exponent of deep time delay reservoirs\label{CLE}}

To use the two-layer deep Ikeda reservoir we enable the input into the
first layer by setting $J_{1}(t)=\kappa_{1}u(t)$ and solve the delay differential equation
system \eqref{ikedaEquations} with $L$ initial history functions
$\bm{v}_{l,0}(s),\ s\in[-\tau_{l},0]$. The fading memory concept \cite{Dambre2012}
states that the reservoir needs to become independent of those history
functions after a certain time. Therefore two identical reservoirs
with different initial conditions need to approximate each other asymptotically.
 From a dynamical perspective, a reservoir has to show generalized
synchronization to its input.

In the following, we check for generalized synchronization of two
unidirectional coupled systems by estimating the maximal conditional Lyapunov exponent of
the driven system. This is done by the auxiliary system method, where
we initialize two identical systems with different initial conditions
and drive both with the same input sequence.
In order to provide comparability to later observations, the input sequence is drawn randomly from an uniform distribution $s(k)\sim \mathcal{U}[-1,1]$, and time multiplexing is used as described in equation \eqref{eq:timemultiplexing} with $T=25$, $N_V=25$, and $\theta=1$.
The conditional Lyapunov exponent then measures
the convergence or divergence rate. If its maximal value is below
zero, the state sequences will approximate each other asymptotically,
and therefore the system shows generalized synchronization to the
input system. If the exponent is positive, the systems diverge.

In order to calculate the conditional Lyapunov exponent  we consider the  distance between the solutions of  two identical systems with different initial conditions:
$\boldsymbol{v}(t)=\boldsymbol{v}(t,\boldsymbol{\phi_{0}})$ and $\boldsymbol{\tilde{v}}(t)=\boldsymbol{v}(t,\boldsymbol{\phi_{0}'})$:
\begin{align}
\dot{\boldsymbol{v}}(t) & =F(\boldsymbol{v}_t,u(t))\nonumber\\
\dot{\boldsymbol{\tilde v}}(t) & =F(\boldsymbol{\tilde v}_t,u(t))\nonumber\\
\dot{\boldsymbol{\mu}}(t) & =\dot{\boldsymbol{\tilde{v}}}(t)-\dot{\boldsymbol{v}}(t)\nonumber\\
 & =F(\boldsymbol{v}_t+\boldsymbol{\mu}_t,u(t)) - F(\boldsymbol{v}_t,u(t)),\label{eq:xidef}
\end{align}
where 
\begin{align}
F(\boldsymbol{v}_t,u(t)) &= 
\begin{bmatrix}
F_1(\boldsymbol{v}_1(t),\boldsymbol{v}_1(t-\tau_1),u(t))\\
F_2(\boldsymbol{v}_2(t),\boldsymbol{v}_2(t-\tau_2),\boldsymbol{v}_1(t))\\
\vdots \\
F_L(\boldsymbol{v}_L(t),\boldsymbol{v}_L(t-\tau_L),\boldsymbol{v}_{L-1}(t))\\
\end{bmatrix}\nonumber\\   
\boldsymbol{v}_t := \boldsymbol{v}(t-s)&=\begin{bmatrix}
\boldsymbol{v}_1(t-s)\\
\boldsymbol{v}_2(t-s)\\
\vdots\\
\boldsymbol{v}_L(t-s)\\
\end{bmatrix},
0 \leq s \leq \tau_\text{max}\label{eq:interval},\\
\tau_\text{max} &= \max_{l=1..L} \tau_l,\nonumber\\
 \boldsymbol{\mu}_t &= \boldsymbol{v}_t - \boldsymbol{\tilde{v}}_t .\nonumber
\end{align}
As a remark, $\boldsymbol{v}_l(t)$ gives the state of layer $l$ whereas $\boldsymbol{v}_t$ is a function of the $L$-layer system state defined over the interval given in \eqref{eq:interval}. The evolution of the distance $\boldsymbol{\mu(t)}$ is now given by a set of delay differential equations. For small
perturbations, we linearise equation \eqref{eq:xidef}: 
\begin{align}
\dot{\boldsymbol{\mu}}(t) & =A{\boldsymbol{\mu}}(t)+\beta B(t){\boldsymbol{\mu}}_t,\label{eq:derXi}
\end{align}
where the linear part of $F$ was summarized into a constant matrix
$A$, and the nonlinearity and the time varying input are included
into $\beta B(t)$.  According to \cite{Hale1993} the solution of
\eqref{eq:derXi} can be estimated as 
\begin{align}
\|\boldsymbol{\mu}(t)\| & \leq e^{\lambda t}\text{const},
\end{align}
where $\lambda$ is the conditional Lyapunov exponent of the non-autonomous reservoir.

For the numerical estimation of the conditional Lyapunov exponent, two equal non-autonomous
systems with different initial conditions of the first layer $x_{1}(t),\ x_{1}^{\prime}(t)$
were evaluated. The input sequence was drawn from the uniform distribution
$s(k)\sim\mathcal{U}[-1,1]$. The distances $\mu_{l}^{num}(t),\ l=1..L$,
of the state sequences was calculated using the maximum norm over
each delay interval $Q(q)=[\left(q-1\right)\tau_{l},q\tau_{l}],\ q\in\mathbb{Z}$:
\begin{equation}
\mu_{l}^{num}(q\tau_{l})=\sup_{t\in Q(q)}\|\boldsymbol{v}_{l}(t)-\boldsymbol{v}_{l}^{\prime}(t)\|.
\end{equation}
An exponential function was approximated accordingly to 
\begin{align}
\mu_{l}^{num}(q\tau_{l})\approx ce^{\lambda_{\max}^{(l)}q\tau_{l}},
\end{align}
where $\lambda_{\max}^{(l)}$ determines the numerically approximated
conditional Lyapunov exponent for layer $l$.

\begin{figure}
\centering \includegraphics[width=1\linewidth]{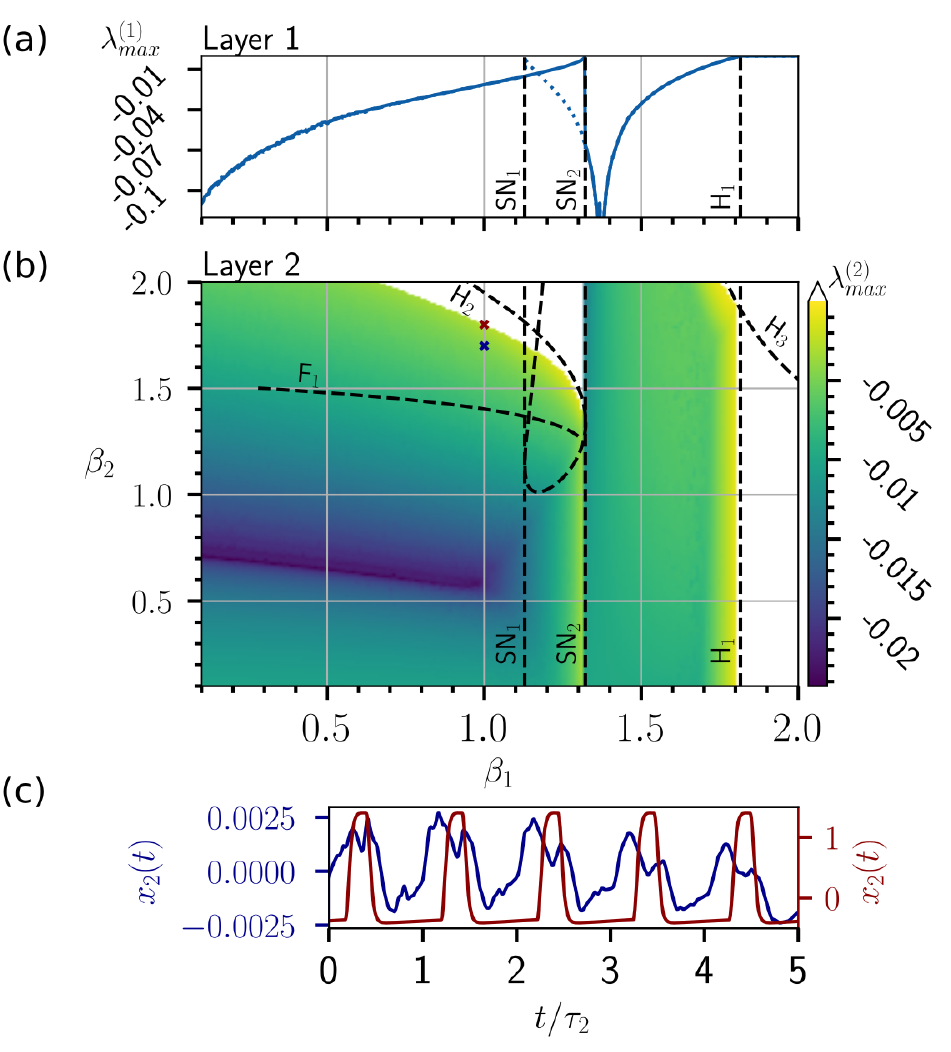}
\caption{\label{lyapExpBetaPlane}Numerical estimations of conditional Lyapunov
exponents of (a) $\lambda_{\max}^{(1)}$ for the first layer as a function of the feedback gain $\beta_1$, and
(b) $\lambda_{\max}^{(2)}$ for the second layer color coded in the parameter plane $(\beta_1,\beta_2)$. Black dashed lines
show the positions of the bifurcations shown in Fig.~\ref{dynamics}.
In panel (a), two lines (solid and dotted) correspond to two
different conditional Lyapunov exponents, which are obtained for two different sets of initial
conditions of the first layer. They appear due to bistability of the
layer dynamics for $\beta_{1}$ between the saddle-node bifurcations.
In (b) the initial conditions are set equally to the calculated conditional Lyapunov exponent corresponding to the solid line in (a).
In (c) the transients of the second layer $x_2(t)$ are shown for feedback gains $\beta_1=1,\beta_2=1.7$ (blue line) and $\beta_1=1,\beta_2=1.8$ (red line) which further correspond to the colored crosses in (b). The latter reveals a periodic oscillation of the length of the delay $\tau_2$, leading to loss of generalized synchronization shown in (b).}
\end{figure}

In Fig.~\ref{lyapExpBetaPlane} the numerical estimation of the
conditional Lyapunov exponent is presented for a two-layer deepTRC. The parameters are as in Table~\ref{tab:Parameter-values}, except the input gain, which was set
to $\kappa_1=0.01$.\footnote{The lower input gain $\kappa_1$ was chosen to create low-degree MC, because
higher degrees are computationally expensive.} Here, the feedback-gains $\beta_{l}$ were scanned systematically for both layers.

In Fig \ref{lyapExpBetaPlane}~(a) the conditional Lyapunov exponent of the first layer
is shown, which  depends only on $\beta_{1}$. Here, the conditional Lyapunov exponent starts
to increase with  increasing $\beta_{1}$ and drops short behind the second 
saddle-node bifurcations SN$_2$ at $\beta_1=1.32$ shown in Fig.~\ref{dynamics}~(a). In the region between the two saddle-node bifurcations SN$_1$ and SN$_2$ according to the bistability of the autonomous system two Lyapunov exponent were computed referring to different initial conditions. The dip at $\beta_{1}\approx 1.38$ ,close after the annihilation of the stable equilibrium $x_{1}^{\ast}=0.1$ with the unstable one, reveals the most negative Lyapunov exponent. After the Hopf bifurcation H$_1$ in the first layer at $\beta_{1}\approx1.8$ the conditional Lyapunov exponent becomes positive, and no generalized synchronization is possible. In other words, the system violates the fading memory condition for $\beta_{1}>1.8$.

The conditional Lyapunov exponent $\lambda_{max}^{(2)}$ computed for the second layer reflects the dynamics of both layers. We observe from Fig.~\ref{lyapExpBetaPlane},
that  $\lambda_{max}^{(2)}$ has, in general, a smaller magnitude
than $\lambda_{max}^{(1)}$. As a result, a perturbation of the first
layer's initial condition stays longer in the system when a second
layer is added. For larger values of $\beta_{1}$, the  borders of
the region, where the conditional Lyapunov exponent is negative, are determined by the Hopf
bifurcations. 
The strong negative conditional Lyapunov exponent in the range $\beta_{1}\in[0.1,1],\beta_{2}\in[0.5,0.75]$ is due to the so-called
exceptional point \cite{Sweeney2019}, were two negative real eigenvalues coalescence.
In the parameter range $\beta_{1}\in[0.7,1.3],\beta_{2}\in[1.5,2]$ the second layer losses generalized synchronization before reaching the subcritical Hopf bifurcation. We assume that due to the ongoing drive of the system, the second layer is pushed into the basin of attracting of the coexisting limit cycle, shown in Fig.~\ref{lyapExpBetaPlane}(c) (red line). Such periodic oscillations lead to a loss of generalized synchronization before reaching the subcritical Hopf H$_2$. Note that the observed oscillations are strongly nonlinear and their shape has a ''switching'' property known for such type of systems \cite{Ruschel2019}.

\section{Conditional Lyapunov exponents versus Memory Capacity}

In this section, we systematically investigate  the relation between
the conditional Lyapunov exponent and the MC for two-layer deepTRC. The MC measures how the reservoir memorizes and transforms previous input \cite{Dambre2012,Koester2020,Harkhoe2019},
for more details about MC and a definition, we refer to Appendix~\ref{app:MC}.
\begin{table}
\begin{tabular}{|c|c|c|}
\hline 
Description & Parameter & Value\tabularnewline
\hline 
\hline 
separation of nodes & $\theta$ & 1\tabularnewline
\hline 
clock cycle & $T$ & 25\tabularnewline
\hline 
virtual nodes per layer & $N_{V}$ & 25\tabularnewline
\hline 
initial steps & $K_{\text{init}}$ & $10^{5}$\tabularnewline
\hline 
training steps &  $K_{\text{train}}$ & $10^{5}$\tabularnewline
\hline 
MC threshold ($N_R=50$) &  $C_{th}$ & $0.012$ \tabularnewline
\hline 
MC threshold ($N_R=75$) &  $C_{th}$ & $0.018$\tabularnewline
\hline
\end{tabular}
\caption{Parameter values used for computation of the memory capacity in Fig.~\ref{betaPlane} and Fig.~\ref{fig:tauDPlane}.\label{tab:McSimParam} }
\end{table}

\begin{figure*}
\includegraphics[width=1\linewidth]{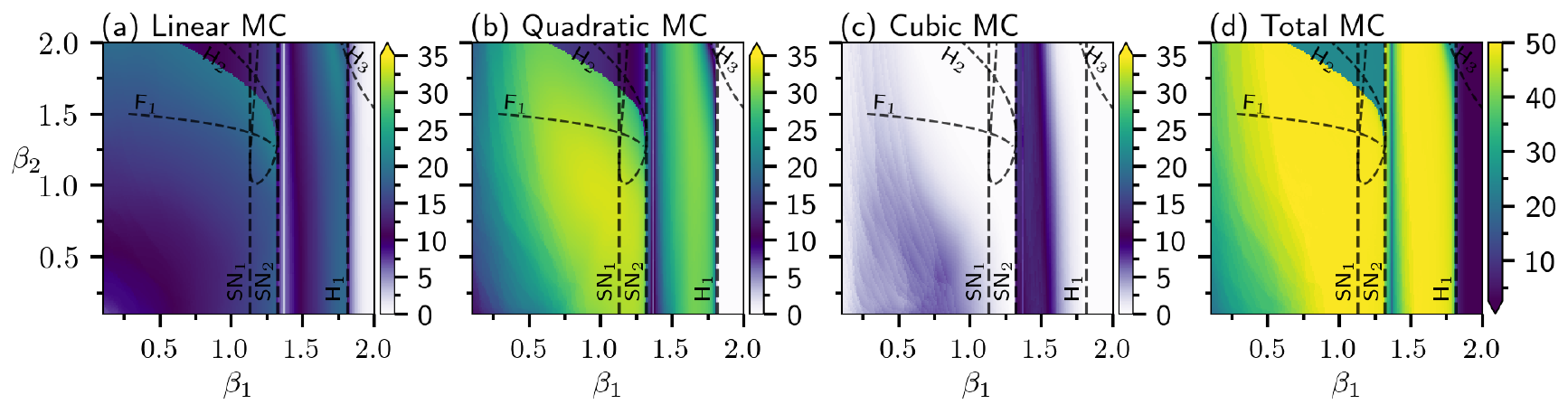} \caption{\label{betaPlane}  Dependence of the memory capacity (MC) on the feedback gains parameters
$\beta_{1}$ and $\beta_{2}$. In (a)-(c) the linear, quadratic and cubic MC is color coded while (d) shows the total MC. The black dashed lines show the positions of the bifurcations shown in Fig.~\ref{dynamics}. The parameters are as given in Table~\ref{tab:Parameter-values}.}
\end{figure*}

In Fig.~\ref{betaPlane}, we show the linear, quadratic, cubic,
and total MC for the parameter values as in Table~\ref{tab:McSimParam}.
The maximal total memory capacity  $MC_{\text{tot}}=50$ can be reached
in a wide parameter range, as shown in Fig.~\ref{betaPlane}~(d),
except for the regions with periodic solutions or strongly negative
 conditional Lyapunov exponent. In all four panels, a clear drop of the MC is visible close
after the supercritical Hopf bifurcation in the first layer due to
the violation of fading memory, i.e., in layer 1 at $\beta_{1}\approx1.816$. 
Also the drop of MC occurs before reaching the subcritical Hopf in layer two, which is in agreement with the loss of generalized synchronization shown in Fig.~\ref{lyapExpBetaPlane}.
More specifically, we observe the following features: 
\begin{itemize}
\item A large memory capacity, which is necessary for a reservoir computer
to perform its tasks, is observed in the regions where conditional Lyapunov exponent is negative,
and the fading memory condition is satisfied.
\item The highest linear MC can be achieved close to the bifurcations where
the  conditional Lyapunov exponent is negative and small in absolute value. In such a case,
the linear information of the input stays longer in the system. 
\item  With the decreasing of the conditional Lyapunov exponent, the linear MC is decreasing, and
MC$_{2}$ starts dominating. With the further decrease of conditional Lyapunov exponent, the
third-order MC becomes dominant. We remark that there is always a
trade-off between the MC of different degrees since the total MC bounds
their sum.
\end{itemize}
Concluding, different dynamical regimes of a deepTRC can boost different
degrees of the MC.

\section{Resonances between Delays and Clock Cycle \label{DelayClockTimeResonances}}

As recently shown analytically and numerically by Stelzer et al.~\cite{Stelzer2019},
resonances between delay-time $\tau$ and clock cycle $T$  lead to
a degradation of linear memory capacity due to parallel alignment
of eigenvectors of the underlying network. This effect was later 
shown in all degrees of the memory capacity by Harkhoe et al.~\cite{Harkhoe2019}
and Köster et al.~\cite{Koester2020} independently. This loss of
total memory capacity at all resonances $a\tau\approx bT,\ a,b\in\mathbb{N}$
of delay-time and clock cycle further results in less performance
at certain tasks.

In the following, we analyze this effect for two and three layer deepTRC
via computing the single degrees of memory capacity up to the cubic
degree while scanning the delays $\tau_{1},\tau_{2}$, and $\tau_{2},\tau_{3}$,
respectively. The system parameters are as in Table~\ref{tab:Parameter-values}
and the simulation parameter are shown in Table~\ref{tab:McSimParam}.

\subsection{Two-Layer deepTRC\label{twoLayerDelay}}

\begin{figure*}
\includegraphics[width=1\linewidth]{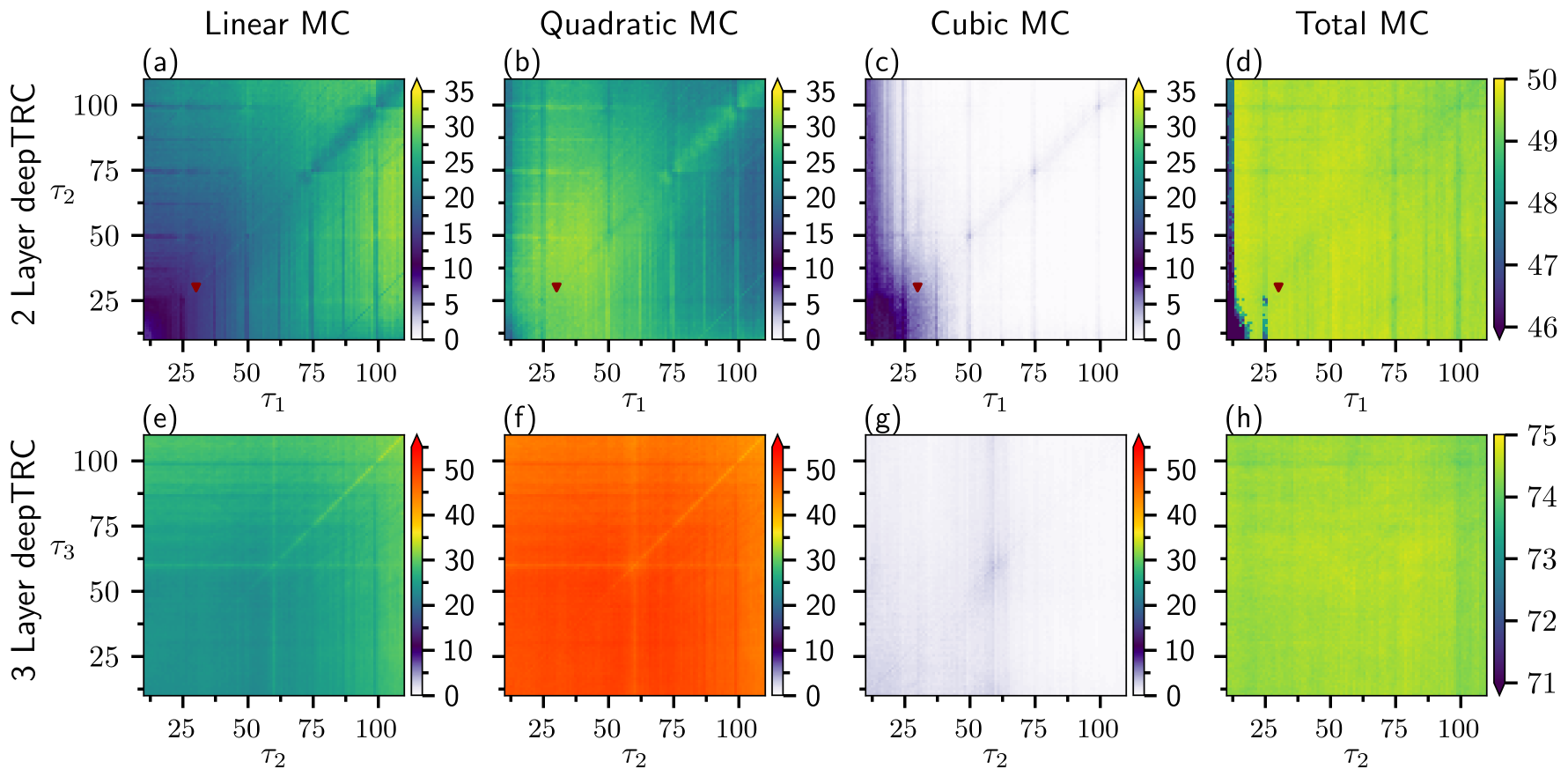}
\caption{\label{fig:tauDPlane}(a) -{}- (c) Memory capacities (MC) of the linear, quadratic, cubic degree and (d) total MC plotted color coded for the two-layer deepTRC. The reservoir size is $N_{R}=50$ and the clock cycle is $T=25$. The red triangles mark the delay setting used in Fig.~\ref{betaPlane}.
(e) -{}- (g) show the linear, quadratic, cubic degree of the MC and (h) the total MC color coded for the three-layer deepTRC with $\tau_{1}=60$ fixed. The reservoir size is $N_{R}=75$
and the clock cycle $T=25$. The feedback gains $\beta_l$ are given in table \ref{tab:Parameter-values}.
The resolution of the delay scan is $\Delta\tau_{i}=0.5$.}
\end{figure*}

In Figs.~\ref{fig:tauDPlane}~(a)–(d) the numerically computed memory capacities are shown as a function of the delays of a two-layer deepTRC. Resonances between the delays
$\tau_{1}$,  $\tau_{2}$, and  the clock cycle $T=25$ are present
in all shown degrees of the MC. Further, resonances of the two delays occur
as diagonals in the plot, with the main diagonal being dominant. In
contrast to the off-diagonal resonances, the $\tau_{1}=\tau_{2}$
resonance broadens with higher delays. The total memory capacity exhibits
weak degradations at the diagonal delay–delay and the clock cycle–delay
resonances. The comparison between the linear and nonlinear MC reveals
the trade-off between both, where the linear MC becomes dominant at
$\tau_{1}\gtrsim75$ and $\tau_{2}<\tau_{1}$.

In contrast to the reported linear MC degradations of the delay-{}-clock
cycle resonances for single-layer TRC, a new effect is visible in
Fig.~\ref{fig:tauDPlane}~(a), where the $\tau_{2}=kT$ resonance
crosses the main diagonal. We observe that for fixed $\tau_{2}=kT$,
when $\tau_{1}$ increases, the linear MC is degraded for $\tau_{1}<\tau_{2}$,
while it is boosted for $\tau_{1}>\tau_{2}$. 

\begin{figure}[hb]
\centering \includegraphics[width=1\linewidth]{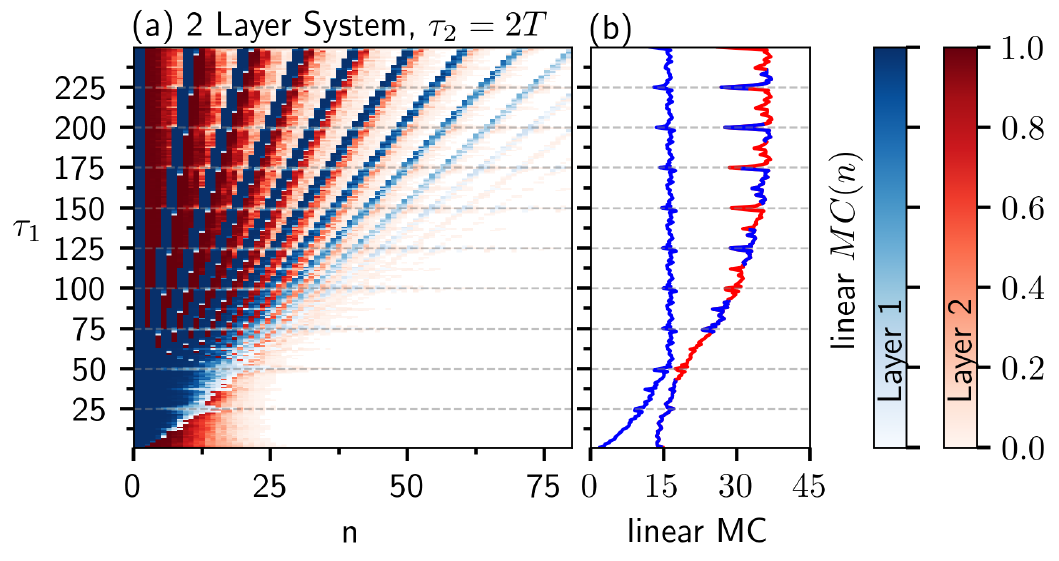}
\caption{\label{fig:mclocs} (a) Linear recallability MC(n) as a function of the recalled past input $n$ and the delay $\tau_1$ depicted as color codes for the single layer system (blue) and an additional second layer (red).
Here, the second layer augments the intervals with low MC of the first
layer. (b) Overall linear memory capacity of the single-layer system (blue) and the
strongly superior two-layer system (red-blue) for increasing delay $\tau_1$. The latter benefits from the appearing augmentation effect.}
\end{figure}

In the following, we present an explanation of a boosted linear MC
of a two-layer deepTRC for $\tau_{1}>\tau_{2}=2T$. For this, we compute
the linear recallability~\cite{Jaeger2004a} $MC(n)\in[0,1]$ of inputs $n$-steps into
the past and separate between the two layers. Hereby, $MC(n)=1$ means the reservoir can fully recover the input from $n$ clock cycles ago. In Fig.~\ref{fig:mclocs}~(a)
the recallability of both layers is presented in different colour. As shown here, increasing $\tau_{1}$ for a single layer TRC leads to an improved recallability from inputs farther in the
past. For comparison, at $\tau_{1}>50$, the single-layer TRC is able
to recall inputs up to $n\approx15$. For $\tau_{1}>2T$, the linear
MC splits up into small intervals with a high MC alternated with intervals
of almost no MC, visible as blue rays in Fig.~\ref{fig:mclocs}.
The length of the intervals of no MC further increases for longer
delay $\tau_{1}$, where the frequency can be estimated as $\tau_{1}/T$.

The addition of a second layer with a delay of $\tau_{2}=2T$, while
the delay of the first layer is $\tau_{1}<2T$, increases the length of
the recallability up to $n=20$. Furthermore, for $\tau_{1}>2T$
the intervals of low MC are augmented by the MC of the second layer. Here, the areas with a high recallability in layer 1 (blue) are augmented
by areas of high recallability in layer 2 (red), which results in
a overall higher linear MC. Accordingly we call the appearing phenomenon augmentation effect. Again, resonance effects at $k\tau_{1}=mT$ occur, resulting in degradations.

\subsection{Three-layer deepTRC}

In Figs.~\ref{fig:tauDPlane}~(e)-(h), we show the MC of a three-layer
deepTRC with a reservoir size of $N_{R}=75$. The delay of the first
layer is  fixed at $\tau_{1}=60$, and the delay plane $(\tau_{2},\tau_{3})$
is scanned. The linear and quadratic MC exhibit strong resonances
at $\tau_{2}=\tau_{1}=60$, $\tau_{3}=\tau_{1}=60$, which increase
the linear MC. The resonances with multiples of the clock cycle occur
as well, but they are less prominent with $\tau_{2},\tau_{3}<\tau_{1}$.
The resonances between $\tau_{2}$ and $\tau_{3}$ occur as the main
diagonal in the plots. 
In comparison to the two-layer case, here the memory is more regular in the shown range $\tau_{2,3}\in[10,120]$. Changes in the memory are more strongly influenced by the first layer delay $\tau_1$. 
The
cubic MC becomes maximal with the delays of all three layers being
equal. The total MC is almost equal for all $\tau_{2}$ and $\tau_{3}$
showing a small general loss and with weak degradations at resonances
with the clock cycle for $\tau_{2},\tau_{3}>\tau_{1}$.

\subsection{Memory Capacity Distribution of deepTRC\label{DeepSpectrum}}

In this section, we show the role of multiple layers of deepTRC and we show ways how to systematically create certain memory in the a deepTRC. 
In particular, we present two configurations: the first one allows using the deep
layers for producing large higher-order MC; the second configuration
produce an increasing in the linear MC with the growing number of
layers.

\begin{figure}[t]
\centering \includegraphics[width=1\linewidth]{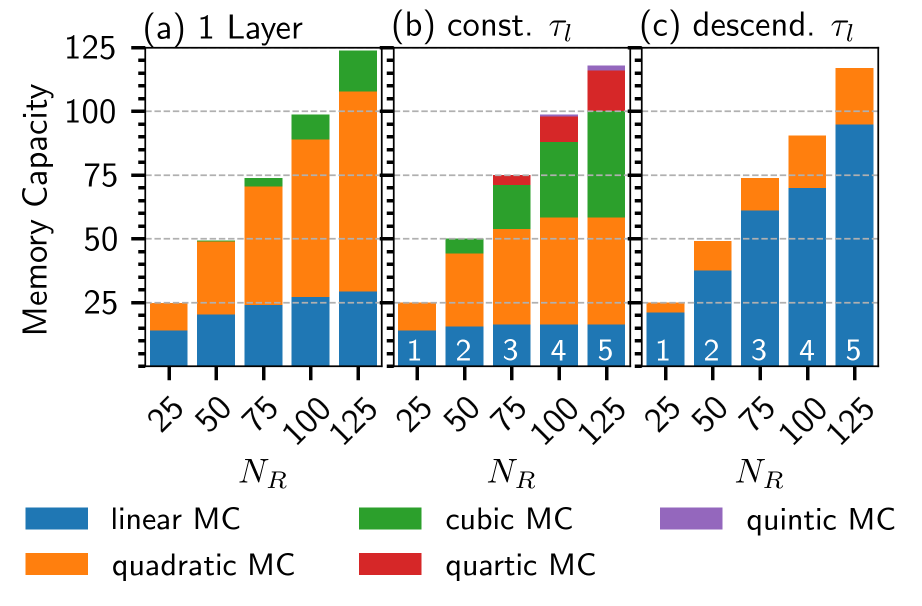}
\caption{\label{deepVsShallow}Distribution of the memory capacities (MC) of different degrees (different colors within the bars) for (a) a single-layer TRC, (b) deepTRC with constant delays $\tau_{l}=27$ and (c) deepTRC with optimal linear MC using descending delays $\tau_{l}=2^{L-l}\cdot60$, e.g. the five layer deepTRC in (c) has the following delays: $\tau_1=960,\tau_2=480,\tau_3=240,\tau_4=120$ and $\tau_5=60$. The white number at the foot of the bar marks the number of layers $L$.}
\end{figure}

We start with a single-layer TRC. Fig.~\ref{deepVsShallow}~(a)
shows the MC distribution with an increasing of the  number of virtual
nodes $N_{V}$. The increase of the nodes number leads mainly to an
increase of quadratic MC, and, starting from $N_{V}=100$, a slow
increase of the cubic MC. As a remark, increasing the virtual nodes
of a single-layer TRC changes the clock cycle and, therefore, we adjusted
the delay to $\tau_{1}=1.2T$. 

In Fig.~\ref{deepVsShallow}~(b), we keep $N_{V}=25$ fixed and
 add layers with the same number of nodes to the deepTRC. The delay of
all layers is kept fixed at $\tau_{l}=27$ and the clock cycle is
$T=25$. We observe that additional layers lead to an increasing of
nonlinear MC of higher orders. 

The third configuration is similar to (b) but now the delays were
varied in order to boost only the linear MC. 
Here, the augmentation effect shown in Fig.~\ref{fig:mclocs} was extended
to $L>2$ layers by using the phenomenologically found rule $\tau_{l}=2^{(L-l)}\cdot60$. 
The last layer has a delay $\tau_{1}=60$ because here a single layer shows its largest
linear MC before splitting into rays. As shown in Fig.~\ref{deepVsShallow}~(c),
this method produces large linear MC, while suppressing higher order MC. 

Possible combinations of the presented delay configurations can
be a good method for adjusting a deepTRCs memory to specific tasks. 
The deepTRC configurations in Fig.~\ref{deepVsShallow} show small general losses of total MC with a increasing number of layers, which is caused by linear dependence of different nodes. 

\section{Conclusion}

We analyzed a Deep Time-Delay Ikeda System and presented a relation
between the dynamics of the autonomous deepTRC and the numerically computed
conditional Lyapunov exponent. We showed a correlation between MC
and conditional Lyapunov exponents. A high linear MC is observed for
small negative conditional Lyapunov exponent. With decreasing of the Lyapunov exponent, different degrees of MC are sequentially activated. Further, we investigated the clock
cycle-delay resonances in different layers as well as delay-delay
resonances. The degradation of the total MC at clock cycle delay
resonances was numerically shown. We explained the boost
of the linear MC at specific delay-clock cycle resonances by an augmentation
effect. Additionally, we used the gained information to present two general
delay configurations, one with an increasing nonlinear MC and one
with boosted linear MC. These configurations provide a variability
superior to a single-layer TRC. They provide a potential
for building general deepTRC, which are oriented at different tasks due to a broad MC spectrum. We could verify that the deepTRC concept
is a promising architecture for fast and variable reservoir computing.

\section*{Acknowledgements}
The authors would like to thank Florian Stelzer for fruitfull discussions.
M.G acknowledges financial support provided by the Deutsche Forschungsgemeinsschaft (DFG, German Research Foundation) through the IRTG 1740. K.L. and F.K. acknowledge support from the Deutsche Forschungsgemeinschaft in the framework of the CRC910. S.Y. acknowledges the financial support by the Deutsche Forschungsgemeinschaft - Project 411803875.

\section*{Data Availability}
The data that support the findings of this study are available from the corresponding author upon reasonable request.

\appendix

\section{Normalised Root Mean Squared Error\label{app:NRMSE}}

The performance at a certain task is measured using the normalized
root mean squared error (NRMSE) given by: 
\begin{equation}
NRMSE=\sqrt{\frac{1}{K_{\text{tst}}}\sum_{k=0}^{K_{\text{tst}}}\frac{\|o(k)-\hat{o}(k)\|}{var(o(k))}},
\end{equation}
where $K_{\text{tst}}$ is the length of the testing sequence, $o(k)$
is the reservoir output and $\hat{o}(k)$ are the task-dependent desired
outputs. Further, $var(\hat{o}(k))$ is the variance of the desired output
and $\|\cdot\|$ is the Euclidean norm.

\section{Stability and Hopf Bifurcation Analysis\label{app:StabilityAnalysis}}

This section gives a detailed analysis of the autonomous dynamics
of the two-layer deep Ikeda time-delay system given in \eqref{ikedaEquations}
which we repeat here for convenience: 
\begin{align}
\dot{x}_{1}(t)= & -x_{1}(t)+\beta_{1}\sin^{2}(x_{1}(t-\tau_{1})+b_{1}),\nonumber \\
\dot{x}_{2}(t)= & -x_{2}(t)-\delta_{2}y_{2}(t)\nonumber \\
 & +\beta_{2}\sin^{2}(x_{2}(t-\tau_{2})+\kappa_{2}x_{1}(t)+b_{2}),\label{eq:2-layers}\\
\dot{y}_{2}(t)= & x_{2}(t).\nonumber 
\end{align}
By setting the derivates to zero, the equations for  the equilibrium
are
given by: 
\begin{align*}
x_{1}^{\ast} & =\beta_{1}\sin^{2}(x_{1}^{*}+b_{1}),\\
x_{2}^{\ast} & =0,\\
y_{2}^{\ast} & =\frac{\beta_{2}}{\delta_{2}}\sin^{2}(\kappa_{2}x_{1}^{*}+b_{2}),
\end{align*}
where the equilibrium $x_{1}^{\ast}$ of the first layer cannot be given
explicitly. In order to check the stability we linearise the system
at its equilibrium leading to: 
\begin{align*}
\dot{\xi}_{1}= & -\xi_{1}(t)+\beta_{1}\sin(2(x_{1}^{\ast}+b_{1}))\xi(t-\tau_{1})\\
\dot{\xi}_{2}= & -\xi_{2}+\beta_{2}\kappa_2\sin(2(\kappa_2 x_{1}^{\ast}+b_{2}))\xi_{1}(t)\\
 & +\beta_{2}\sin(2(\kappa_2 x_{1}^{\ast}+b_{2}))\xi_{2}(t-\tau_{2})-\delta\eta_{2}(t)\\
\dot{\eta_{2}}= & \xi_{2}(t)
\end{align*}
This can be rewritten into a lower triangular block matrix form as shown in section \ref{Dynamics} and
therefore the resulting characteristic equation is given by:
\begin{align*}
0 & =\underbrace{(-\lambda-1+\beta_{1}\nu_{1}e^{-\lambda\tau_{1}})}_{\text{Layer 1}}\cdot\underbrace{(\delta+\lambda(\lambda+1-\beta_{2}\nu_{2})e^{-\lambda\tau_{2}}))}_{\text{Layer 2}}
\end{align*}
with $\nu_{1}=\sin(2(x_{1}+b))$ and $\nu_{2}=\sin(2(\kappa x_{1}+b_{1})$
and lambda being the eigenvalues of the autonomous deepTRC.

An equilibrium $(x_{1}^{\ast},x_{2}^{\ast},y_{2}^{\ast})$ is asymptotically
stable if the real part of every solution of the characteristic equation
is negative: 
\begin{equation}
Re(\lambda_{i})<0\ \forall i,
\end{equation}
what we check for in the following for layer 1 and layer 2 separately.

\subsection*{Stability of Layer 1}

The eigenvalues of the first layer are roots of the equation $-\lambda-1+\beta_{1}\nu_{1}e^{-\lambda\tau_{1}}=0$,
and they are given by the Lambert-$W$ function: 
\begin{align*}
\lambda_{i} & =\frac{1}{\tau_{1}}W_{i}(\beta_{1}\nu_{1}\tau_{1}e^{\tau_{1}})-1
\end{align*}
with $W_{i}(\dots)$ being the $i$-th order of the Lambert-$W$ function.


\subsubsection*{}

The stability of the equilibrium point will change if the eigenvalues
cross the imaginary axis. Therefore we set $\lambda=i\omega$ 
\begin{align*}
i\omega=-1+\beta_{1}\nu_{1}e^{i\phi},
\end{align*}
where  $\phi=\omega\tau_{1}$. Separating the real and imaginary part
we can rewrite this into: 
\begin{align*}
1=\beta_{1}\nu_{1}\cos(\phi),\\
\omega=\beta_{1}\nu_{1}\sin(\phi),\\
\end{align*}
using the absolute values this leads to: 
\begin{align*}
\omega_{H}\tau_{1}^{\ast}=\phi^{\ast}=\arccos(\frac{1}{\beta_{1}\nu_{1}}),\\
\omega_{H}=\sqrt{\beta_{1}^{2}\nu_{1}^{2}-1},\\
\tau_{1}^{\ast}=\dfrac{\arccos(\frac{1}{\beta_{1}\nu_{1}})}{\sqrt{\beta_{1}^{2}\nu_{1}^{2}-1}},
\end{align*}
with $\tau_{1}\leq\tau_{1}^{\ast}$ being stable and $\tau_{1}>\tau_{1}^{\ast}$
being unstable due to a Andronov-Hopf bifurcation.

\subsection*{Stability of Layer 2}

For the second layer we analyse the second term of the characteristic
equation given by: 
\begin{align}
\delta+\lambda(\lambda+1-\beta_{2}\nu_{2})e^{\lambda\tau_{2}}=0.
\end{align}

By substituting $\lambda=i\omega,$ it is straightforward to find the
frequencies, at which the equation can cross the imaginary axis 
\begin{equation}
\bar{\omega}^{2}=-\frac{1-2\delta-(\beta_{2}\nu_{2})^{2}}{2}\pm\frac{\sqrt{(1-2\delta-(\beta_{2}\nu_{2})^{2})^{2}-\delta^{2}}}{4}.
\end{equation}
The Hopf Bifurcation of the second layer occur in regions where: $\bar{\omega}_{\pm}>0$.
Denoting:
\begin{align*}
\phi(\omega)=\arctan2(\sign(\beta_{2}\nu_{2})\omega,\delta-\omega^{2}),\\
\phi(\omega_{-})=\omega_{-}\tau_{H}+2\pi k,\ k\in\mathbb{Z}.
\end{align*}
the delay values for stabilizing $\tau_{H,k}-$ and destabilizing
$\tau_{H,k}+$ Hopf bifurcations are given as: 
\begin{align}
\tau_{H,k}-=-\frac{\phi(\omega_{-})}{\omega_{-}}+\frac{2\pi k}{\omega_{-}},\\
\tau_{H,k}+=-\frac{\phi(\omega_{+})}{\omega_{+}}+\frac{2\pi k}{\omega_{+}}.
\end{align}

\section{Memory Capacity\label{app:MC}}

The task-independent memory capacity (MC) introduced by Dambré et
al.~\cite{Dambre2012} determines how a dynamical system memorizes
previous inputs and how it further transforms them. The total MC is
given by the sum over all degrees $d=1,\dots,\infty,\ d\in\mathbb{N}$
of MC. 
\begin{equation}
MC_{\text{tot}}=\sum_{d=1}^{D=\infty}MC_{d}
\end{equation}
In the following $d=1$ refers to the linear and $d>1$ to the nonlinear
MC (quadratic, cubic, …). Hereby, the linear MC is given by a simple
linear recall of past inputs, whereas the nonlinear gives evidence
about which computation of past inputs are performed by the system.
Further, it was proven that the read-out dimension of a system bounds
the maximal reachable total MC  $MC_{\text{tot}}\leq N_{R}$. According
to this fact, a trade-off between the linear and nonlinear MC can
be obtained.

The MC can be calculated via the correlation between input and the
reservoir states: 
\begin{align}
C[x,\hat{o}] & =\dfrac{\langle\hat{o}X\rangle_{T}\langle X^{T}X\rangle_{K}^{-1}\langle X^{T}\hat{o}\rangle_{K}}{\langle\hat{o}^{2}\rangle_{K}},\\
\langle\hat{o}\rangle_{K} & =\dfrac{1}{K}\sum_{t=1}^{K}\hat{o}(t)\label{eq:MCcorr}
\end{align}
where $\langle\dots\rangle_{K}$ is the average value over the time
$K$, $^{-1}$ is the inverse and $^{T}$ the transpose of the matrix.
The calculation of the MCs via the correlation given in \eqref{eq:MCcorr}
is biased due to statistics and this bias strongly depends on the
length of $K$. Therefore we manually set a threshold meaning that
no MCs below this threshold are regarded.\\
 As suggested by Dambre et al.\cite{Dambre2012}, the input values
$s(k)$ were drawn from a uniform distribution $s\sim\mathcal{U}[-1,1]$.
In order to compute the different degrees of MC we used the set of
Legendre polynomials $\mathcal{L}_{p}$, which provide orthogonality
over the given input range.

For example, a target of the cubic degree $d=3$ and three variables
is given by: 
\begin{align}
\hat{o}(k)= & \mathcal{L}_{1}(k-n_{1})\mathcal{L}_{1}(k-n_{2})\mathcal{L}_{1}(k-n_{3}),\\
 & n_{1}<n_{2}<n_{3}.\nonumber 
\end{align}
In order to find all appearing $C_{d}(\dots)$ a maximal step into
the past of $n_{max}=1000$ was set.

\bibliographystyle{apsrev}
\bibliography{ReservoirComputing.bib}

\end{document}